\documentclass[aps,twocolumn,prd,amsmath,nofootinbib,preprintnumbers,superscriptaddress]{revtex4-1}
\usepackage{epsfig,slashed,nicefrac}
\usepackage[utf8]{inputenc}

\newcommand{\amc}{{\sc MadGraph5\textunderscore}a{\sc MC@NLO}}
\newcommand{\fr}{{\sc Feyn\-Rules}}

\newcommand{\ml}{{\sc MadLoop}}
\newcommand{\mfks}{{\sc MadFKS}}
\newcommand{\nloct}{{\sc NLOCT}}

\newcommand{\sq}{\tilde q}
\newcommand{\st}{\tilde t}
\newcommand{\go}{\tilde g}
\newcommand{\epsbar}{\bar\epsilon}
\newcommand{\bof}{B_0}
\newcommand{\buf}{B_1}
\newcommand{\bofp}{B_0^\prime}
\newcommand{\bufp}{B_1^\prime}

\def\be{\begin{equation*}}
\def\ee{\end{equation*}}
\def\bsp#1\esp{\begin{split}#1\end{split}} 
\def\bpm{\begin{pmatrix}}
\def\epm{\end{pmatrix}}

\begin{document}

\preprint{CERN-PH-TH-2015-224,  DCPT/15/118, IPPP/15/59, LPSC/15/267, MCNET-15-27, SLAC-PUB-16400}

\title{Matching next-to-leading order predictions to parton showers in
  supersymmetric QCD}

\author{C\'eline Degrande}
\affiliation{
  Institute for Particle Physics Phenomenology,
  Department of Physics Durham University, Durham DH1 3LE,
  United Kingdom}
\author{Benjamin Fuks}
\affiliation{Sorbonne Universit\'es, UPMC Univ.~Paris 06, UMR 7589, LPTHE, F-75005 Paris, France}
\affiliation{CNRS, UMR 7589, LPTHE, F-75005 Paris, France}
\author{Valentin Hirschi}
\affiliation{SLAC, National Accelerator Laboratory,
  2575 Sand Hill Road, Menlo Park, CA 94025-7090, USA}
\author{Josselin Proudom}
\affiliation{Laboratoire de Physique Subatomique et de Cosmologie,
  Universit\'e Grenoble-Alpes, CNRS/IN2P3, 53 Avenue des Martyrs, F-38026
  Grenoble Cedex, France}
\author{Hua-Sheng Shao}
\affiliation{CERN, PH-TH, CH-1211 Geneva 23, Switzerland}

\begin{abstract}
  We present a fully automated framework based on the \fr\ and \amc\ programs
  that allows for accurate simulations of supersymmetric QCD processes at the
  LHC. Starting directly from a model Lagrangian that features squark and gluino
  interactions, event generation is achieved at the next-to-leading order in
  QCD, matching short-distance events to parton showers and including the
  subsequent decay of the produced supersymmetric particles. As an application,
  we study the impact of higher-order corrections in gluino pair-production in a
  simplified benchmark scenario inspired by current gluino LHC searches.
\end{abstract}

\maketitle

\textit{Introduction} --
The LHC has been designed with the aim of exploring the electroweak symmetry
breaking mechanism and to possibly shed light on phenomena beyond the Standard
Model. During its first run, both the ATLAS and CMS collaborations have
extensively investigated many different channels in order to get hints for new
physics. However, no striking signal has been
found so that limits have been set on popular models, such as the Minimal
Supersymmetric Standard Model (MSSM)~\cite{Nilles:1983ge,Haber:1984rc},
or on simplified models for new physics~\cite{Alwall:2008ag,Alves:2011wf}. All
these searches will nevertheless be pursued during the next LHC runs,
benefiting from larger statistics and higher center-of-mass energy. In this
work, we focus on MSSM-inspired simplified new physics scenarios in which gluino
pairs can be copiously produced at the LHC. As in many related experimental
searches~\cite{Aad:2014wea,Aad:2015iea,Chatrchyan:2014lfa,Khachatryan:2015vra},
we consider the case where both produced gluinos decay into a pair of jets
and an invisible neutralino, and then revisit the phenomenology of such models.

Experimental gluino analyses are currently based on Monte Carlo simulations of
the signals where leading-order (LO) matrix elements of different partonic
multiplicities are matched to parton showers and merged. The predictions are
then normalized to resummed total rates combined with the next-to-leading order
(NLO) result~\cite{Beenakker:1996ch,Kulesza:2008jb,Kulesza:2009kq,%
Beenakker:2009ha,GoncalvesNetto:2012yt}.
More sophisticated differential theoretical predictions are however always
helpful for setting more accurate exclusion limits, possibly refining the search
strategies, and measuring the model free parameters in case of a
discovery~\cite{Dreiner:2010gv}. In this context, it has been recently
demonstrated that the \amc\ framework~\cite{Alwall:2014hca} can provide a
general platform for computing (differential) observables within many beyond the
Standard Model theories~\cite{Degrande:2014sta}. This approach relies in
particular on the use of the \fr~\cite{Alloul:2013bka} and \nloct~\cite{%
Degrande:2014vpa} packages for (automatically) generating a UFO library~\cite{%
Degrande:2011ua} containing the vertices and the needed counterterms for NLO
computations.

Within this framework, we match for the first time NLO QCD matrix-element-%
based predictions to parton showers for gluino pair-production. Virtual
contributions are evaluated following the Ossola-Papadopoulos-Pittau (OPP)
formalism as implemented in \ml~\cite{Ossola:2006us,Ossola:2007ax,%
Hirschi:2011pa} and combined
with real emission contributions by means of the FKS subtraction method as
embedded in \mfks~\cite{Frixione:1995ms,Frederix:2009yq}; these two
modules being fully incorporated in \amc. The
matching to parton showers is then achieved by employing the MC@NLO method~\cite{%
Frixione:2002ik}. After accounting for (LO) gluino decays, we study the impact
of both the NLO contributions and the parton showers in the context of LHC
physics.

\medskip

\textit{Theoretical framework} --
Our study of gluino pair-production and decay is based on an MSSM-inspired
simplified model. We complement the Standard Model with three
generations of non-mixing left-handed and right-handed squark fields $\sq_{L,R}$
of mass $m_{\tilde q_{L,R}}$ and with a Majorana fermionic gluino field $\go$
of mass $m_{\tilde g}$. The dynamics of the new fields is described by the
following Lagrangian,
\be\bsp
 & {\cal L}_{\rm SQCD} =
      D_\mu \sq_L^\dag D^\mu \sq_L
    + D_\mu \sq_R^\dag D^\mu \sq_R
    + \frac{i}{2}\bar\go\slashed{D}\go\\
  &\quad
    - m_{\sq_L}^2 \sq_L^\dag \sq_L
    - m_{\sq_R}^2 \sq_R^\dag \sq_R
    - \frac12m_{\go}\bar\go\go\\
  &\quad
    + \sqrt{2} g_s \Big[ - \tilde q_L^\dag T \big(\bar{\tilde g} P_L q \big)
    + \big(\bar q P_L \tilde g \big) T \tilde q_R  + {\rm h.c.} \Big] \\
  &\quad
   - \frac{g_s^2}{2}
     \Big[\tilde q_R^\dag T \tilde q_R - \tilde q_L^\dag T \tilde q_L\Big]
     \Big[\tilde q_R^\dag T \tilde q_R - \tilde q_L^\dag T \tilde q_L\Big] \ ,
\esp \ee
that contains all interactions allowed by QCD gauge invariance and
supersymmetry, as well as squark and gluino kinetic and mass terms. In our
notation, $T$ stands for the fundamental representation matrices of $SU(3)$,
$P_L$ ($P_R$) for the left-handed (right-handed) chirality projector and $g_s$
is the strong coupling constant. Flavor and color indices are left
understood for brevity.

In addition, we enable the (possibly three-body) decays of the colored
superpartners by including (s)quark couplings to a gauge-singlet Majorana
fermion $\chi$ of mass $m_\chi$ that is identified with a bino,
\be\bsp
 & {\cal L}_{\rm decay} =
    \frac{i}{2}\bar\chi\slashed{\partial}\chi - \frac12m_\chi\bar\chi\chi\\
 &\quad
   + \sqrt{2} g' \Big[ -\tilde q_L^\dag Y_q \big(\bar\chi P_L q \big)
     + \big(\bar q P_L \chi \big) Y_q \tilde q_R  + {\rm h.c.} \Big]\ .
\esp \ee
In this Lagrangian, $Y_q$ denotes the hypercharge quantum number of the
(s)quarks and $g'$ the hypercharge coupling.

\begin{table*}[!t]
\renewcommand{\arraystretch}{1.25}
\setlength{\tabcolsep}{12pt}
 \begin{tabular}{c||cc}
    $m_{\tilde{g}}$~[GeV] & $\sigma^{\rm LO}$~[pb] & $\sigma^{\rm NLO}$~[pb] \\
    \hline \hline
   200 & $2104^{+30.3\%}_{-21.9\%}{}^{+14.0\%}_{-14.0\%}$                  &
         $3183^{+10.8\%}_{-11.6\%} {}^{+1.8\%}_{-1.8\%}$ \\
   500 & $15.46^{+34.7\%}_{-24.1\%}{}^{+19.5\%}_{-19.5\%}$                 &
         $24.90^{+12.5\%}_{-13.4\%}{}^{+3.7\%}_{-3.7\%}$ \\
   750 & $1.206^{+35.9\%}_{-24.6\%}{}^{+23.5\%}_{-23.5\%}$                 &
         $2.009^{+13.5\%}_{-14.1\%}{}^{+5.5\%}_{-5.5\%}$\\
  1000 & $1.608\cdot 10^{-1}{}^{+36.3\%}_{-24.8\%}{}^{+26.4\%}_{-26.4\%}$ &
         $2.743\cdot 10^{-1}{}^{+14.4\%}_{-14.8\%}{}^{+7.3\%}_{-7.3\%}$\\
  1500 & $6.264\cdot 10^{-3}{}^{+36.2\%}_{-24.7\%}{}^{+29.4\%}_{-29.4\%}$ &
         $1.056\cdot 10^{-2}{}^{+16.1\%}_{-15.8\%}{}^{+11.3\%}_{-11.3\%}$ \\
  2000 & $4.217\cdot 10^{-4}{}^{+35.6\%}_{-24.5\%}{}^{+29.8\%}_{-29.8\%}$ &
         $6.327\cdot 10^{-4}{}^{+17.7\%}_{-16.6\%}{}^{+17.8\%}_{-17.8\%}$ \\
\end{tabular}
\renewcommand{\arraystretch}{1.0}
\caption{\small \label{tab:totalxs}LO and NLO QCD inclusive cross sections for
  gluino pair-production at the LHC, running at a center-of-mass energy of
  $\sqrt{s}=13$~TeV. The results are shown together with the associated scale
  and PDF relative uncertainties.}
\end{table*}

At the NLO in QCD, gluino pair-production receives contributions from real
emission diagrams as well as from the
interferences of tree-level diagrams with virtual one-loop diagrams that exhibit
ultraviolet divergences. These must be absorbed through a suitable
renormalization of the parameters and of the fields appearing in
${\cal L}_{\rm SQCD}$. To this aim, we replace all (non-)fermionic
bare fields $\Psi$ ($\Phi$) and bare parameters $y$ by the corresponding
renormalized quantities,
\be\bsp
  &\Phi \to \big[1 + \frac12 \delta Z_\Phi\big] \Phi\ ,\ \
  \Psi \to \big[ 1 + \frac12 \delta Z^L_\Psi P_L +
      \frac12 \delta Z^R_\Psi P_R \big]\Psi\ ,\\
  & \hspace{3cm} y \to y + \delta y \ ,
\esp\ee
where the renormalization constants $\delta Z$ and $\delta y$ are truncated at
the first order in the strong coupling $\alpha_s$.

The wave-function renormalization constants of the massless quarks
($\delta Z_q^{L,R}$), of the top quark ($\delta Z_t^{L,R}$), of the gluon
($\delta Z_g$) and the top mass renormalization constant
($\delta m_t$) are given, when adopting the on-shell renormalization scheme, by
\be\bsp
  & \delta Z_g = -\frac{g_s^2}{24 \pi^2} \bigg[
    - \frac13
    + \bof(0,m_t^2,m_t^2)
    + 2 m_t^2 \bofp(0,m_t^2,m_t^2)
   \\&\quad
    - \frac{n_c}{3}
    + n_c\bof(0,m_{\go}^2,m_{\go}^2)
    + 2 n_c m_{\go}^2 \bofp(0,m_{\go}^2,m_{\go}^2)
   \\&\quad
    +\sum_{\sq}\Big[\frac16 + \frac14 \bof(0,m_{\sq}^2,m_{\sq}^2)
        - m_{\sq}^2 \bofp(0,m_{\sq}^2,m_{\sq}^2)\Big]
    \bigg]\ , \\
  & \delta Z_q^{L,R} = \frac{g_s^2 C_F}{8 \pi^2}\
     \buf(0; m_{\go}^2, m_{\sq_{L,R}}^2) \ , \\
  & \delta Z_t^{L,R} = \frac{g_s^2 C_F}{16 \pi^2} \bigg[
        1 + 2 \buf(m_t^2;m_{\go}^2,m_{\st_{L,R}}^2)
  \\ &\quad
      + 2 \buf(m_t^2;m_t^2,0)
      + 8 m_t^2 \bofp(m_t^2; m_t^2,0)
  \\ &\quad
      + 4 m_t^2 \bufp(m_t^2; m_t^2,0)
      + 2 m_t^2 \sum_{i=L,R}\bufp(m_t^2;m_{\go}^2,m_{\st_i}^2)\bigg] \ , \\
  & \delta m_t = -\frac{g_s^2 C_F m_t}{16 \pi^2} \bigg[
      -\! 1 \!+\! 4 \bof(m_t^2; m_t^2,0)
       \!+\! 2 \buf(m_t^2; m_t^2,0)
 \\ & \quad
       + \sum_{i=L,R}\buf(m_t^2;m_{\go}^2,m_{\st_i}^2) \bigg] \ ,
\esp\ee
where the $B_{0,1}$ (and $A_0$, for further references) functions and their
derivatives stand for the standard two-point (one-point) Passarino-Veltman
loop-integrals~\cite{Passarino:1978jh}. Moreover, \mbox{$n_c=3$} and
\mbox{$C_F=(n_c^2-1)/(2 n_c)$} denote respectively the number of colors and the
quadratic Casimir invariant associated with the fundamental representation of
$SU(3)$. The gluino wave-function and mass renormalization constants
$\delta Z_{\go}^{L,R}$ and $\delta m_{\go}$ are given by
\be\bsp
  & \delta Z_{\go} = \frac{g_s^2}{16\pi^2} \bigg[
        n_c
      + 2 n_c \buf(m_{\go}^2;m_{\go}^2,0)
  \\&\quad
      + 8 n_c m_{\go}^2 \bofp(m_{\go}^2;m_{\go}^2,0)
      + 4 n_c m_{\go}^2 \bufp(m_{\go}^2;m_{\go}^2,0)
  \\&\quad
      + \sum_{\sq=\sq_L,\sq_R}\Big\{
          \buf(m_{\go}^2;m_q^2,m_{\sq}^2) +
          2 m_{\go}^2 \bufp(m_{\go}^2;m_q^2,m_{\sq}^2)\Big\}
   \bigg]\ ,\\
  & \delta m_{\go}   = \frac{g_s^2 m_{\go}}{16\pi^2} \bigg[
        n_c
      - 4 n_c \bof(m_{\go}^2;m_{\go}^2,0)
      - 2 n_c \buf(m_{\go}^2;m_{\go}^2,0)
  \\&\quad
      - \sum_{\sq=\sq_L,\sq_R} \buf(m_{\go}^2;m_q^2,m_{\sq}^2)
   \bigg]\ ,
\esp\ee
while the squark wave-function ($\delta Z_{\sq}$) and mass ($\delta m_{\sq}^2$)
renormalization constants read,
\be\bsp
  & \delta Z_{\sq} =  \frac{g_s^2 C_F}{8\pi^2} \bigg[
     - \bof(m_{\sq}^2;m_{\go}^2,m_q^2)
     + \bof(m_{\sq}^2;m_{\sq}^2,0)
  \\&\ \
     \!+\! (m_{\go}^2\!-\!m_{\sq}^2\!+\!m_q^2) \bofp(m_{\sq}^2;m_{\go}^2,m_q^2)
     \!+\! 2 m_{\sq}^2 \bofp(m_{\sq}^2;m_{\sq}^2,0)
   \bigg] \ , \\
  & \delta m_{\sq}^2 = \frac{g_s^2 C_F}{8\pi^2} \bigg[
       A_0(m_{\sq}^2) - A_0(m_{\go}^2)
     - 2 m_{\sq}^2 \bof(m_{\sq}^2;m_{\sq}^2,0)
  \\&\ \
     + (m_{\sq}^2-m_{\go}^2-m_q^2) \bof(m_{\sq}^2;m_{\go}^2,m_q^2)
        - A_0(m_q^2)
   \bigg] \ ,\\
\esp\ee
with \mbox{$(-)^L\equiv1$} and \mbox{$(-)^R\equiv-1$}, and with
\mbox{$m_q\neq0$} for top squarks only. As a result
of the structure of the gluino-squark-quark interactions, squark mixing effects
proportional to the corresponding quark masses are generated at the one-loop
level and must be accounted for in the renormalization procedure. In our
simplified setup, we consider $n_f=5$ flavors of massless quarks so that these
effects are only relevant for the sector of the top squarks. In this case,
matrix renormalization is in order,
\be
  \bpm \st_L\\ \st_R\epm \to
  \bpm \st_L\\ \st_R\epm + \frac12 \bpm
   \delta Z_{\st_L}       & \delta Z_{\st, \rm LR} \\
   \delta Z_{\st, \rm RL} & \delta Z_{\st_R} \epm
  \bpm \st_L\\ \st_R\epm \ ,
\ee
and we impose that the stop sector is renormalized so that left-handed and
right-handed stops are still defined as non-mixed states at the
one-loop level. In the MSSM, this is made possible by stop couplings to the
Higgs sector that generate an off-diagonal mass counterterm,
\be
  \delta{\cal L}_{\rm off} = - \delta m^2_{\st, \rm LR}
     (\st_L^\dag \st_R + \st_R^\dag \st_L) \ .
\ee
These Higgs couplings being absent in our simplified model, we therefore
introduce $\delta{\cal L}_{\rm off}$  explicitly. The
off-diagonal stop wave-function (\mbox{$\delta Z_{\st, {\rm LR}}=
\delta Z_{\st, {\rm RL}}$}) and mass ($\delta m^2_{\st, \rm LR}$)
renormalization constants are then found to be
\be\bsp
  & \delta Z_{\st, {\rm LR}} =
    \frac{g_s^2 C_F m_{\go} m_t}{4\pi^2(m_{\st_R}^2-m_{\st_L}^2)}
       \sum_{i=L,R} (-)^i \bof(m_{\st_i}^2;m_t^2,m_{\go}^2)\ , \\
  & \delta m^2_{\st, \rm LR} = \frac{g_s^2 C_F m_{\go} m_t}{8\pi^2}
       \sum_{i=L,R} \bof(m_{\st_i}^2;m_t^2,m_{\go}^2) \ ,
\esp\ee
where $\delta Z_{\st, {\rm LR}}$ has been symmetrized. In this way, it
incorporates
the renormalization of the stop mixing angle (taken vanishing in our model)
which does not need to be explicitly introduced~\cite{Eberl:2001eu}.

In order to ensure that the running of $\alpha_s$ solely originates from gluons
and $n_f$ active flavors of light quarks, we renormalize the strong coupling by
subtracting at zero-momentum transfer, in the gluon self-energy, all massive
particle contributions. This gives
\be\bsp
  & \frac{\delta\alpha_s}{\alpha_s} =
      \frac{\alpha_s}{2\pi\bar\epsilon}
        \bigg[\frac{n_f}{3} - \frac{11 n_c}{6}\bigg]
    + \frac{\alpha_s}{6\pi}
        \bigg[\frac{1}{\bar\epsilon} - \log\frac{m_t^2}{\mu_R^2}\bigg]
\\ &\quad 
    + \frac{\alpha_s n_c}{6\pi}
        \bigg[\frac{1}{\bar\epsilon} - \log\frac{m_{\go}^2}{\mu_R^2}\bigg]
    + \frac{\alpha_s}{24\pi}\sum_{\sq}
        \bigg[\frac{1}{\bar\epsilon} - \log\frac{m_{\sq}^2}{\mu_R^2}\bigg]\ .
\esp\ee
The ultraviolet-divergent parts of $\delta\alpha_s/\alpha_s$
are written in terms of the quantity \mbox{$\frac{1}{\epsbar}=\frac{1}{\epsilon} -
\gamma_E + \log{4\pi}$} where $\gamma_E$ is the Euler-Mascheroni constant and
$\epsilon$ is connected to the number of space-time dimensions $D=4-2\epsilon$.

Finally, the artificial breaking of supersymmetry by the mismatch of the two
gluino and the ($D-2$) transverse gluon degrees of freedom must be compensated
by finite counterterms. Imposing that the definition of the strong coupling
$g_s$ is identical to the Standard Model one, only quark-squark-gluino vertices
and four-scalar interactions have to be shifted~\cite{Martin:1993yx},
\be\bsp
&  {\cal L}_{\rm SCT} = 
    \sqrt{2} g_s\frac{\alpha_s}{3 \pi}
      \Big[ - \tilde q_L^\dag T_a \big(\bar{\tilde g}^a P_L q \big)
         + \big(\bar q P_L \tilde g^a \big) T_a \tilde q_R 
         + {\rm h.c.} \Big] \\
 &\quad
     + \frac{g_s^2}{2}\frac{\alpha_s}{4 \pi}
        \Big[ \sq_R^\dag\{T_a,T_b\}\sq_R + \sq_L^\dag\{T_a,T_b\}\sq_L\Big]
\\ &\qquad\qquad\times
        \Big[ \sq_R^\dag\{T^a,T^b\}\sq_R + \sq_L^\dag\{T^a,T^b\}\sq_L\Big]\\
  &\quad
   - \frac{g_s^2}{2}\frac{\alpha_s}{4\pi}
     \Big[\tilde q_R^\dag T_a \tilde q_R - \tilde q_L^\dag T_a \tilde q_L\Big]
     \Big[\tilde q_R^\dag T^a \tilde q_R - \tilde q_L^\dag T^a \tilde q_L\Big] \ ,
\esp\ee
where we have introduced adjoint color indices for clarity.

In our phenomenological study, loop-calculations are performed numerically in
four dimensions by means of the \ml\ package and therefore require the
extraction of rational parts that are related to the $\epsilon$-pieces of the
loop-integral denominators ($R_1$, which are automatically reconstructed within
the OPP reduction procedure) and numerators ($R_2$). For any renormalizable
theory, the number of $R_2$ terms is finite and they can be seen as counterterms
derived from the bare Lagrangian~\cite{Ossola:2008xq}. In the context of the
${\cal L}_{\rm SQCD}$ Lagrangian, all necessary $R_2$ counterterms can be found
in Ref.~\cite{Shao:2012ja}.

The setup described above has been implemented in the \fr\ package and we have
made use of the \nloct\ program to automatically calculate all the ultraviolet
and $R_2$ counterterms of the model. The specificity of the renormalization of
the stop sector has been implemented via a new option of \nloct,
\texttt{SupersymmetryScheme->"OS"}, that allows to treat all scalar fields that
mix at the loop-level as described above. We have validated the output against
our analytical calculations, and these results represent the first validation
of \nloct\ in the context of computations involving massive Majorana colored
particles. We have finally generated a UFO version of the model that can be
loaded into \amc\ and which we have made publicly available on
\verb+http://feynrules.irmp.ucl.ac.be/wiki/NLOModels+.
We have further validated the model, together with the numerical treatment of
the loop-diagrams by \ml, by comparing \amc\ predictions to those of the code
{\sc Prospino}~\cite{Beenakker:1997ut}, using a fully degenerate mass spectrum
due to the limitations of the latter.

\medskip

\begin{figure*}
\centering
  \includegraphics[width=.85\columnwidth]{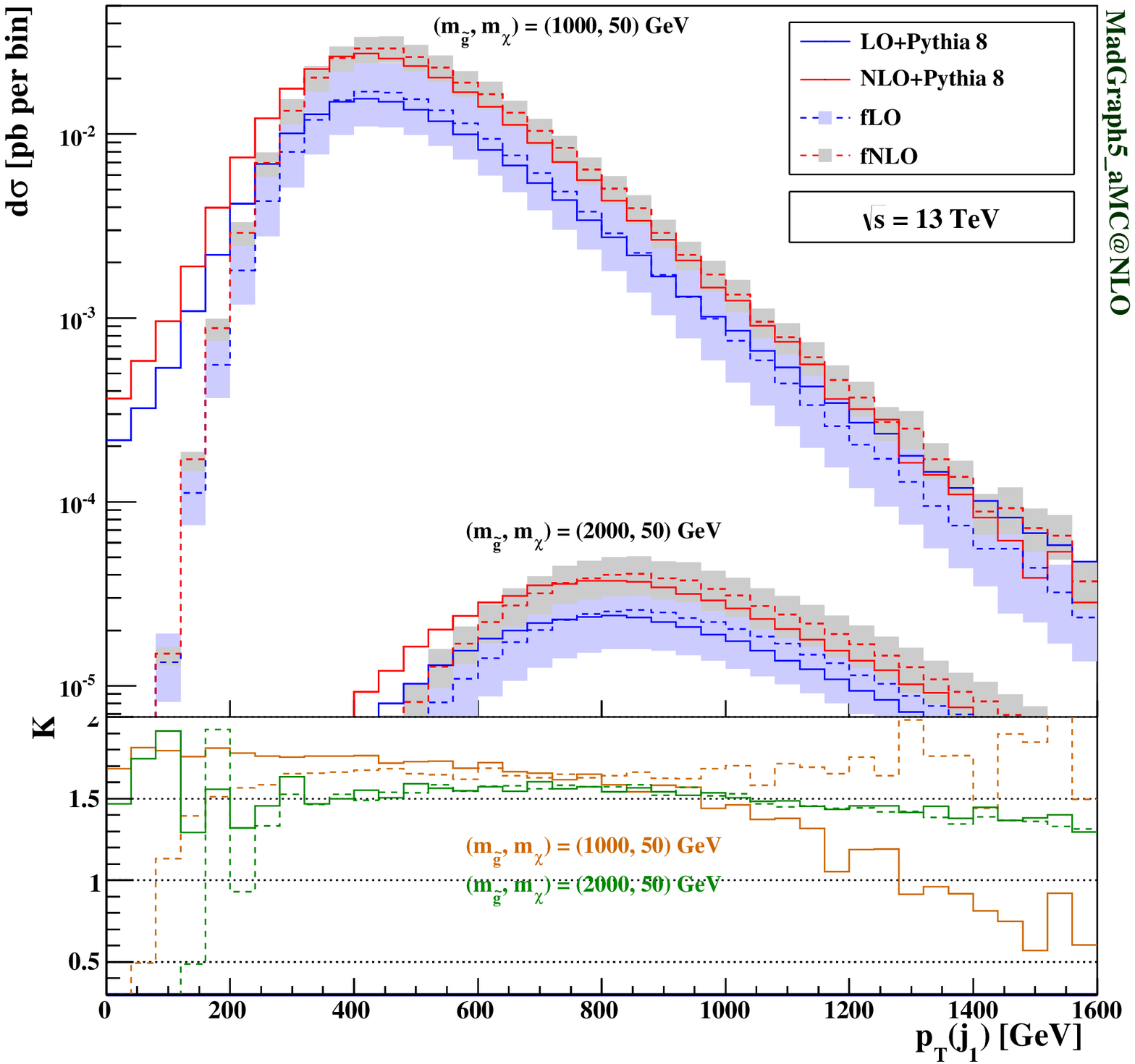}\qquad\qquad
  \includegraphics[width=.85\columnwidth]{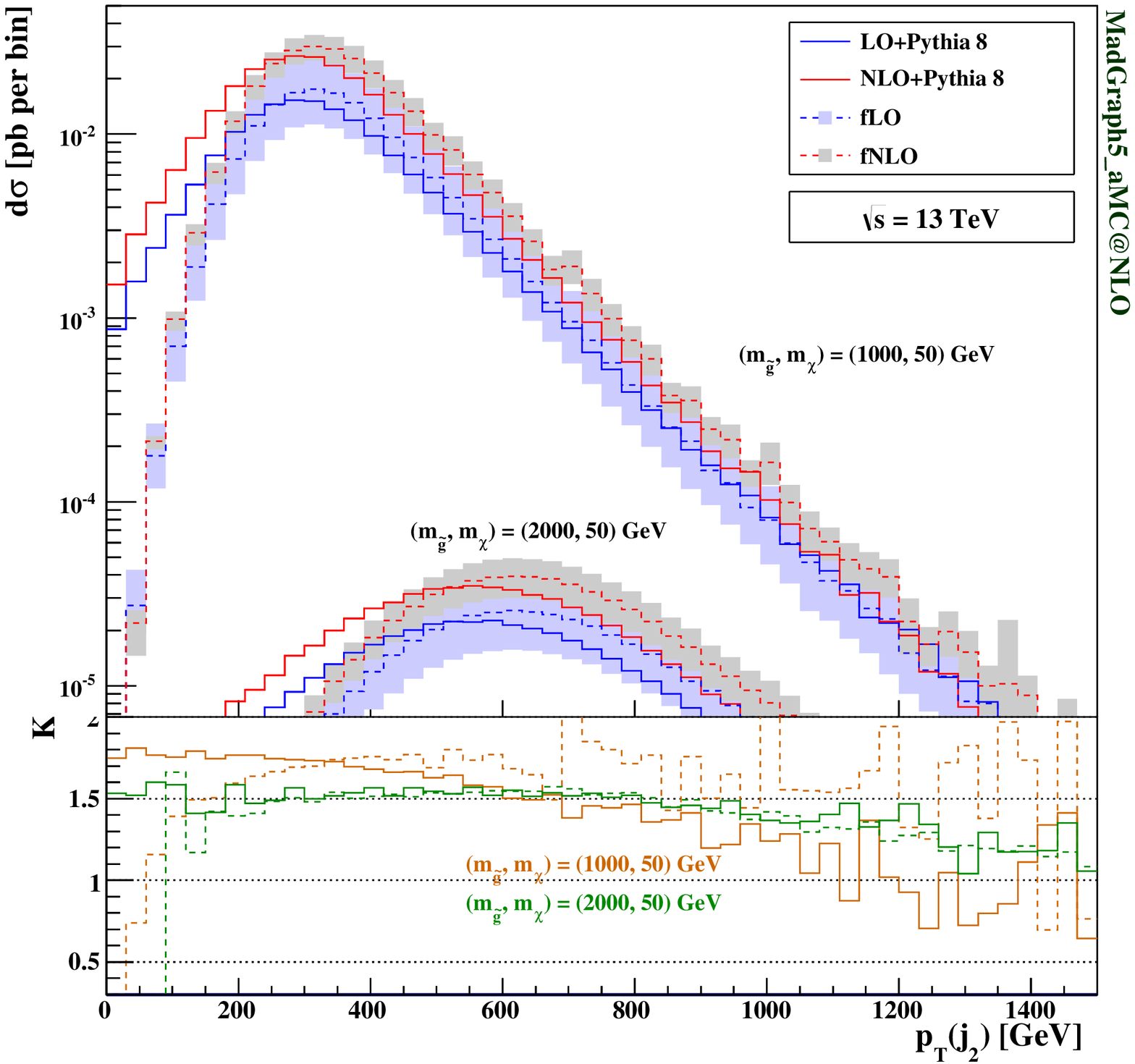}\\
  \includegraphics[width=.85\columnwidth]{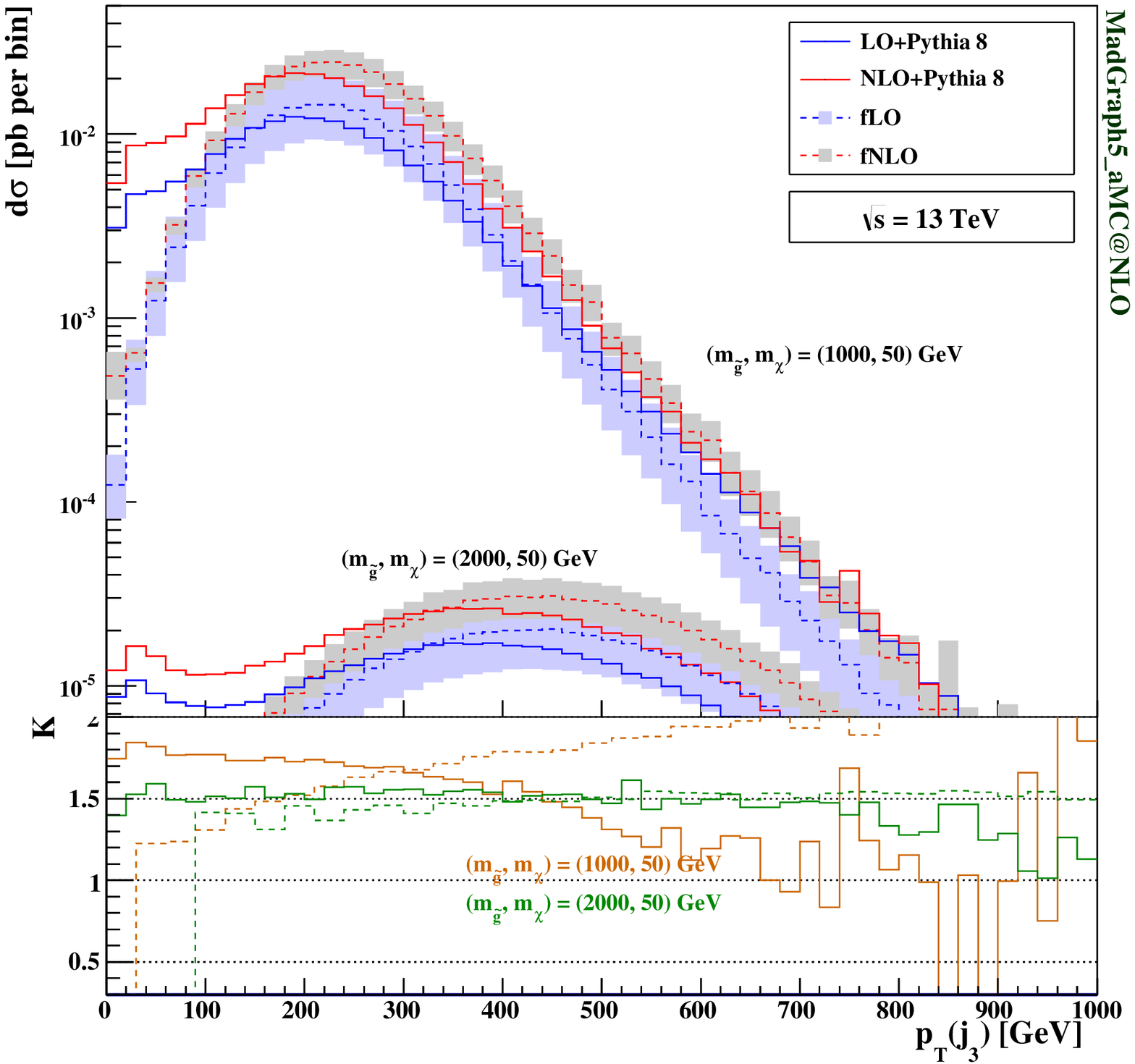}\qquad\qquad
  \includegraphics[width=.85\columnwidth]{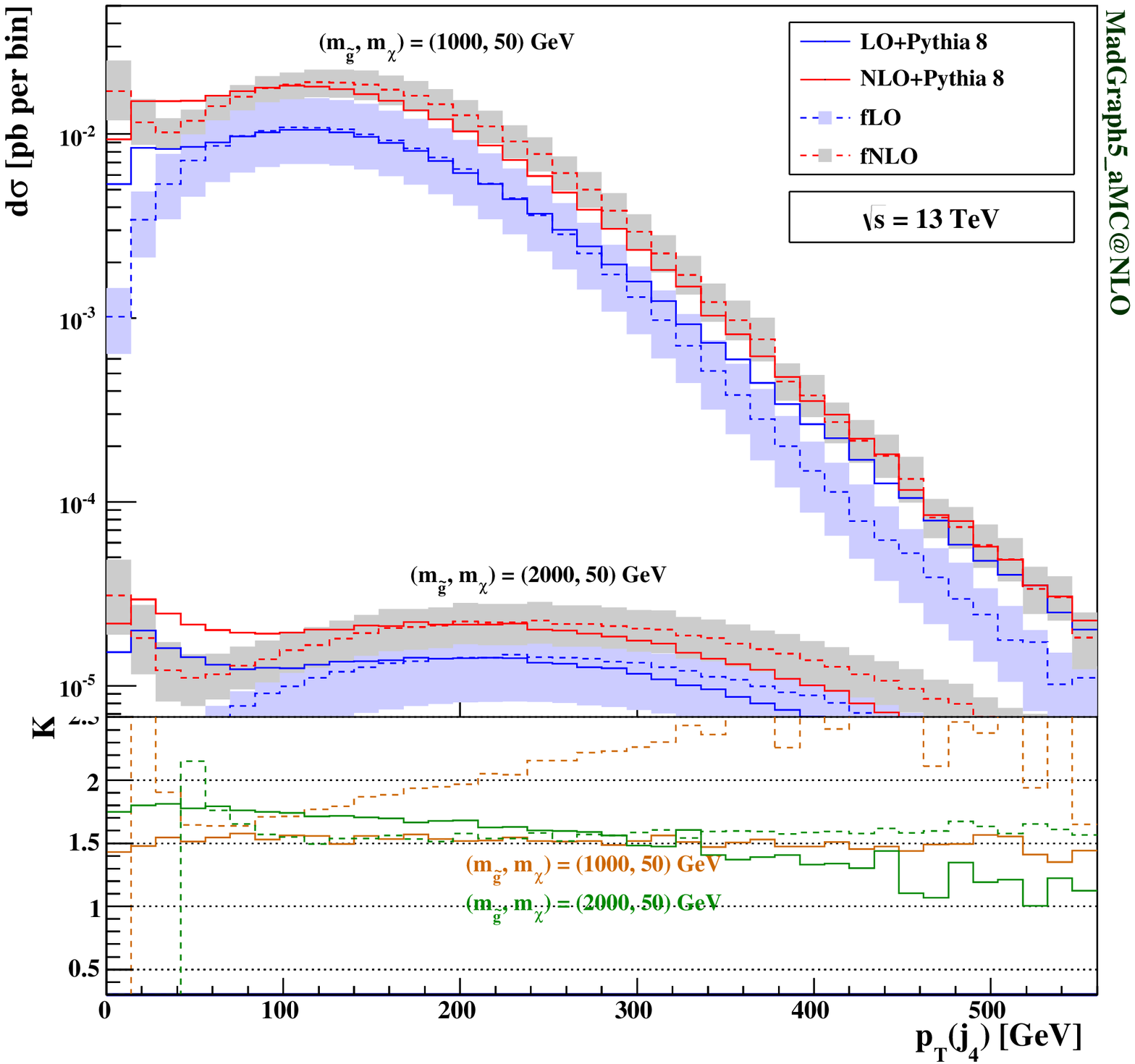}\\
  \includegraphics[width=.85\columnwidth]{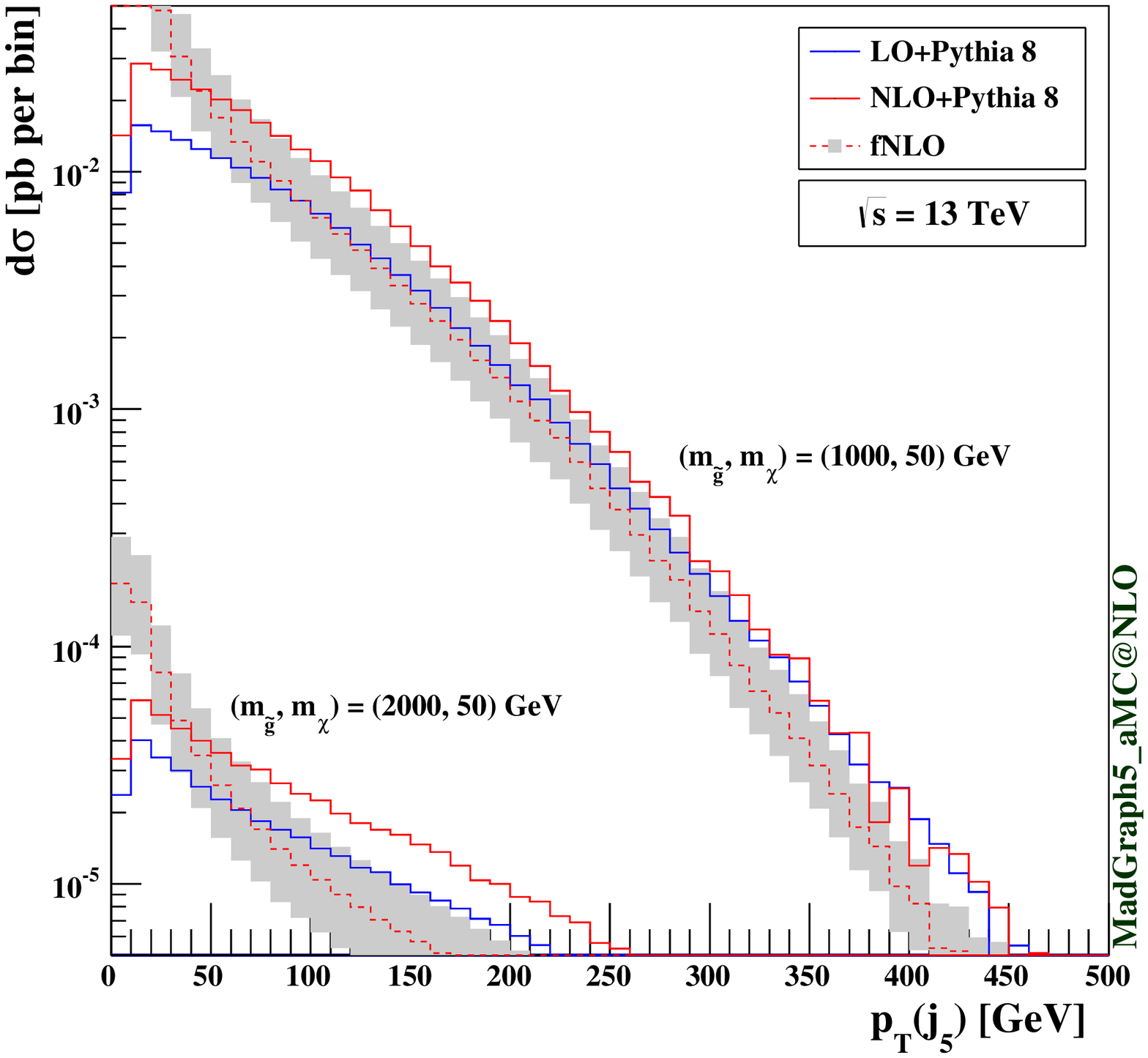}\qquad\qquad
  \includegraphics[width=.85\columnwidth]{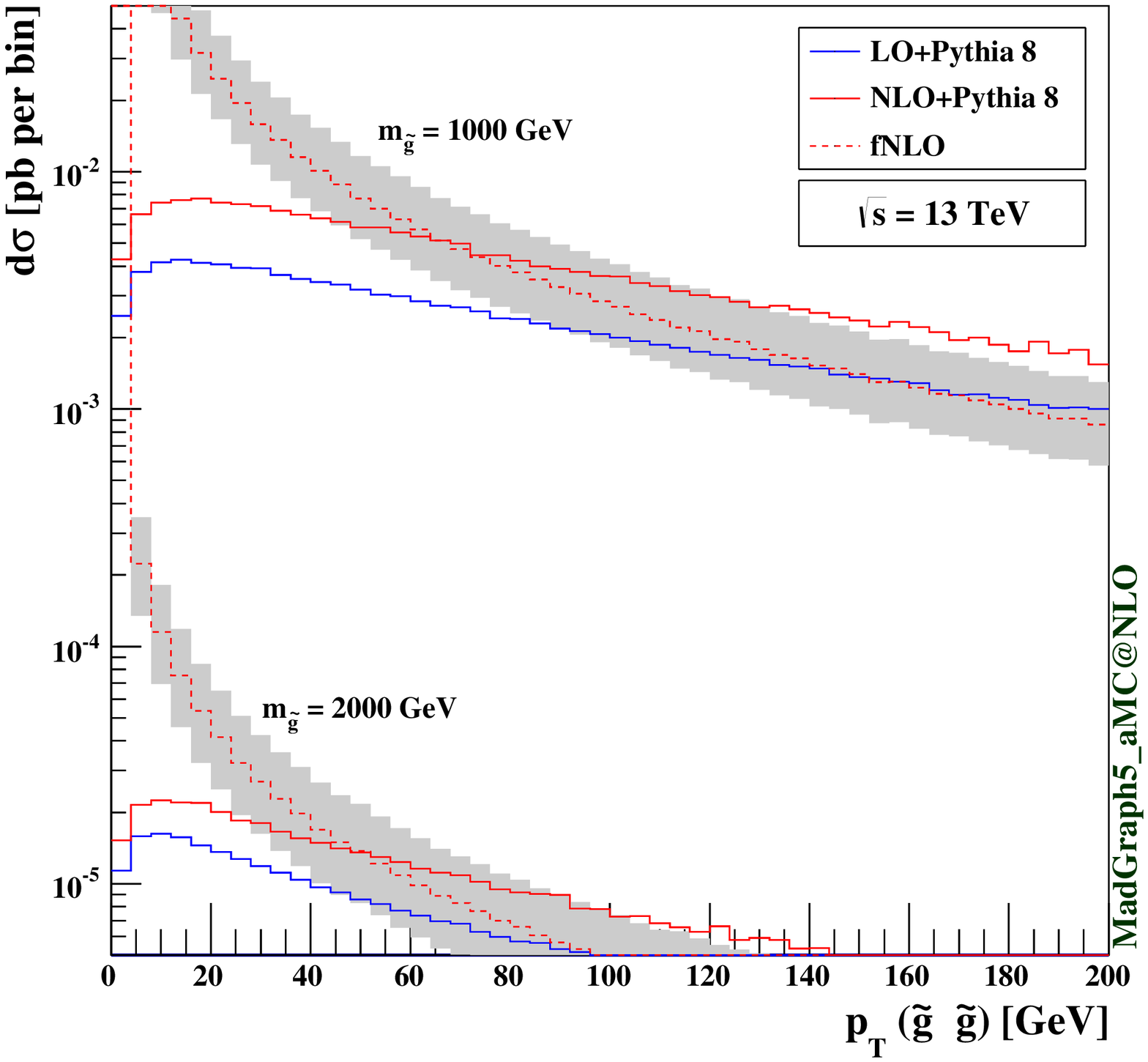}
  \caption{\small \label{fig:ptj}First five leading jet and gluino-pair
   transverse-momentum spectra for the production of a pair of gluinos decaying
   each into two colored partons and a neutralino. We consider two mass
   configurations and
   show results at the NLO (red) and LO (blue) accuracy in QCD, at the
   fixed-order (dashed, fLO and fNLO) and after matching to the {\sc Pythia}~8
   parton shower description (solid). Theoretical
   uncertainties related to the fixed order calculations are shown as blue (LO)
   and gray (NLO) bands. The lower insets of the figure present ratios
   of NLO results to LO ones, both at fixed order (dashed) and after matching
   to parton showers (solid).}
\end{figure*}

\textit{LHC phenomenology} --
In Table~\ref{tab:totalxs}, we compute gluino pair-production total cross
sections for proton-proton collisions at a center-of-mass energy of
\mbox{$\sqrt{s}=$13~TeV} and for different gluino masses. Squarks are
decoupled (\mbox{$m_{\tilde t_L}=16$~TeV}, \mbox{$m_{\tilde t_R} =17$~TeV}
and \mbox{$m_{\tilde q_L} = m_{\tilde q_R}=15$~TeV}) so that any resonant
squark contribution appearing in the real-emission topologies is off-shell and
therefore suppressed. The latter production modes can be seen as the associated
production of a gluino and a squark that subsequently decays into a gluino and a
quark. Including these contributions as parts of the NLO QCD corrections for
gluino pair-production would hence result in a double-counting when considering
together all superpartner production processes inclusively. Moreover, these
resonant channels require a special treatment in the fully-automated \amc\
framework, that is left to future work~\cite{ToAppear}. Our choice for the
squark spectrum corresponds to the one made by ATLAS and CMS collaborations in
their respective gluino searches~\cite{Aad:2014wea,Aad:2015iea,%
Chatrchyan:2014lfa,Khachatryan:2015vra}.

Our results are evaluated both at the LO and NLO accuracy in QCD and presented
together with scale and parton distribution (PDF) uncertainties. For the central
values, we set the renormalization and factorization scales to half the sum of
the transverse mass of all final state particles and use the NNPDF~3.0 set of
parton distributions~\cite{Ball:2014uwa}
accessed via the LHAPDF~6 library~\cite{Buckley:2014ana}.
Scale uncertainties are derived by varying both scales independently by factors
$\nicefrac{1}{2}$, $1$ and $2$, and the PDF uncertainties have been extracted
from the cross section values spanned by all NNPDF distribution replicas following
the NNPDF recommendations~\cite{Demartin:2010er}. We observe a significant
enhancement of the cross section of about 50\% due to genuine NLO contributions,
as well as a sizable reduction of the uncertainties. In particular, the apparent
drastic reduction of the PDF uncertainties is related to the poor quality of the
LO NNPDF fit when compared to the NLO fit~\cite{Ball:2014uwa}.

In order to achieve realistic simulations of LHC collisions, we first handle
gluino decays into two colored partons and a neutralino via an off-shell squark
by using tree-level decay matrix-elements. For the fixed-order results presented below,
we have analytically calculated those decay matrix elements and integrated them over
the phase space, after checking our results with the
{\sc MadSpin}~\cite{Artoisenet:2012st} and
{\sc MadWidth}~\cite{Alwall:2014bza} programs. These latter two programs
have been used in the computation of the NLO predictions matched to parton showers.
Due to the three-body nature of
the gluino decays, spin correlations are here omitted as {\sc MadSpin} can
only handle them for two-body decays. We then interface the partonic events
obtained in this way to a parton showering and hadronization description as
provided by the {\sc Pythia}~8 package~\cite{Sjostrand:2014zea}, and use the
anti-$k_T$ jet reconstruction algorithm~\cite{Cacciari:2008gp} with a radius
parameter set to 0.4, as implemented in {\sc FastJet}~\cite{%
Cacciari:2011ma}, to reconstruct all final state parton-level and
hadron-level jets for fixed-order and parton-shower-matched calculations
respectively. Finally, the phenomenological analysis of the generated events is
performed with {\sc MadAnalysis}~5~\cite{Conte:2012fm}.

Key differential distributions particularly sensitive to both NLO and
shower effects are presented in Fig.~\ref{fig:ptj}. We show the
transverse-momentum ($p_T$) spectra of the first five leading jets
(first five subfigures) for two benchmark
scenarios featuring either a light (\mbox{$m_{\go}=1$~TeV}) or a heavy
(\mbox{$m_{\go}=2$~TeV}) gluino, as well as a rather light bino
(\mbox{$m_\chi=50$~GeV}). We compare fixed-order predictions (dashed) to results
matched to parton showers (solid) and consider both LO (blue) and NLO (red)
accuracy in QCD. We observe that most of the differential $K$-factors (\textit{i.e.},
the bin-by-bin ratios of the NLO result to the LO one) both with and without
parton-shower matching strongly depend on the jet $p_T$ in the considered $p_T$ range. The NLO effects
therefore not only increase the overall normalization of the distributions, but
also distort their shapes. The $K$-factor is indeed greater at low $p_T$
than at high $p_T$, so that the traditional procedure
of using LO predictions scaled by inclusive $K$-factors cannot be used for an
accurate gluino signal description.

Fig.~\ref{fig:ptj} also underlines the effects of matching LO and NLO matrix
elements to parton showers. Since most parton-level jets originate from the decay
of very massive gluinos, the fixed-order $p_T$ distributions peak at large
$p_T$ values. In addition, the low-$p_T$ region of these spectra is
depleted, with the exception of the fourth and fifth jet $p_T$
spectra where radiation effects are non-negligible. As a result of the matching to parton showers, the fixed-order
NLO distributions are distorted and softened. While the change is milder in the
large-$p_T$ tails whose shapes are controlled by the hard matrix element, the
low-$p_T$ regions are mostly sensitive to effects due to multiple emissions and
hence become populated. The
parton shower emissions from hard partons are indeed often not reclustered back
with the jet they are issued from, hence distorting the jet $p_T$ spectra.
For this reason, resummation effects become significant below the peak of the
various $p_T$ distributions. This effect is also illustrated on the last
subfigure of Fig.~\ref{fig:ptj}, where we show
the $p_T$ spectrum of the gluino pair in the small $p_T$ range. We have verified
that the matched results agree with the fixed-order ones for very large $p_T$
values of the order of the gluino mass.

\medskip

\textit{Conclusions} --
We have performed the first calculation of NLO supersymmetric QCD
corrections to gluino pair-%
production matched to parton showers and have studied the impact of both the NLO
contributions and of the parton showers. We have shown that observable effects
could be induced on quantities typically used to obtain exclusion limits and that more accurate calculations are
crucial for extracting model parameters in case of a discovery.

Our calculation has been performed fully automatically and we have applied it
to the case of a simplified model similar to one of those used by the
ATLAS and CMS collaborations for their respective gluino searches. In addition,
we have publicly released the UFO model associated with our computation, that is
sufficiently general to be readily used to explore the phenomenology associated
with any supersymmetric QCD process.

Finally, our results also shows that all technical obstacles for automating the
matching of fixed-order calculations for inclusive supersymmetric particle
production at the NLO in QCD to parton showers have been cleared, up to the
ambiguity issue of the double counting arising in real emission resonant
contributions~\cite{ToAppear}.

\acknowledgments
We are grateful to R.~Frederix, S.~Frixione, F.~Maltoni and O.~Mattelaer
for enlightening discussions.
This work has been supported in part by the ERC grant 291377 \textit{LHCtheory:
Theoretical predictions and analyses of LHC physics: advancing the precision
frontier}, the Research Executive Agency of the European Union under Grant
Agreement PITN-GA-2012-315877 (MCNet) and the Theory-LHC-France initiative of
the CNRS (INP/IN2P3). CD is a Durham International Junior Research Fellow,
VH is supported by the SNF grant PBELP2 146525 and
JP by a PhD grant of the \textit{Investissements d'avenir, Labex ENIGMASS}.

\bibliography{biblio}

\end{document}